\documentclass[sn-mathphys,Numbered]{sn-jnl}% Math and Physical Sciences Reference Style
\usepackage{graphicx}%
\usepackage{multirow}%
\usepackage{amsmath,amssymb,amsfonts}%
\usepackage{amsthm}%
\usepackage{mathrsfs}%
\usepackage[title]{appendix}%
\usepackage{xcolor}%
\usepackage{textcomp}%
\usepackage{manyfoot}%
\usepackage{booktabs}%
\usepackage{algorithm}%
\usepackage{algorithmicx}%
\usepackage{algpseudocode}%
\usepackage{listings}%
\usepackage{placeins}
\usepackage{natbib}

\theoremstyle{thmstyleone}%
\theoremstyle{thmstyletwo}%

\theoremstyle{thmstylethree}%

\begin{document}

\title[Article Title]{Analysis of Research Trends in Computer Science: A Network Approach}

%%=============================================================%%
%% Prefix	-> \pfx{Dr}
%% GivenName	-> \fnm{Joergen W.}
%% Particle	-> \spfx{van der} -> surname prefix
%% FamilyName	-> \sur{Ploeg}
%% Suffix	-> \sfx{IV}
%% NatureName	-> \tanm{Poet Laureate} -> Title after name
%% Degrees	-> \dgr{MSc, PhD}
%% \author*[1,2]{\pfx{Dr} \fnm{Joergen W.} \spfx{van der} \sur{Ploeg} \sfx{IV} \tanm{Poet Laureate} 
%%                 \dgr{MSc, PhD}}\email{iauthor@gmail.com}
%%=============================================================%%

\author*[1]{\fnm{Ghazal} \sur{Kalhor}}\email{kalhor.ghazal@ut.ac.ir}

\author[2]{\fnm{Behnam} \sur{Bahrak}}\email{b.bahrak@teias.institute}

\affil*[1]{\orgdiv{School of Electrical and Computer Engineering, College of Engineering}, \orgname{University of Tehran}, \orgaddress{ \city{Tehran}, \country{Iran}}}

\affil*[2]{\orgname{Tehran Institute for Advanced Studies}, \orgaddress{ \city{Tehran}, \country{Iran}}}

%%==================================%%
%% sample for unstructured abstract %%
%%==================================%%

\abstract{Nowadays, computer science (CS) has emerged as a dominant force in numerous research areas both within and beyond its own discipline. However, despite its significant impact on scholarly space, only a limited number of studies have been conducted to analyze the research trends and relationships within computer science. In this study, we collected information on fields and subfields from over 2,000 research articles published in the 2022 proceedings of the top Association for Computing Machinery (ACM) conferences spanning various research fields. Through a network approach, we investigated the interconnections between CS fields and subfields to evaluate their interdisciplinarity and multidisciplinarity. Our findings indicate that computing methodologies and privacy and security stand out as the most interdisciplinary fields, while human-centered computing exhibits the highest frequency among the papers. Furthermore, we discovered that machine learning emerges as the most interdisciplinary and multidisciplinary subfield within computer science. These results offer valuable insights for universities seeking to foster interdisciplinary research opportunities for their students.}


%%\pacs[JEL Classification]{D8, H51}

%%\pacs[MSC Classification]{35A01, 65L10, 65L12, 65L20, 65L70}

\maketitle

\section{Introduction}\label{sec1}

In recent decades, computer science has emerged as a prominent field of interest among scholars worldwide \cite{kalhor2022new}. Numerous papers have been published across various domains of computer science. Notably, interdisciplinary research papers that employ computer science methodologies to address research inquiries in diverse fields such as economics, biology, and sociology are being accepted not only in computer science journals and conferences but also in top-tier publication venues of other disciplines \cite{ruiz2021artificial, tadmor2005interdisciplinary, lazer2009computational}. Therefore, many researchers from other disciplines collaborate with computer scientists on interdisciplinary projects to keep up with the latest changes in technology.

Several previous studies have analyzed networks of science to assess the relationship between different fields of science in scholarly space. Karunan et al. \cite{karunan2017discovering} explored the interdisciplinarity of the relationship between biotechnology for energy and nanotechnology for energy fields. They utilized citation networks to conduct this study. In \cite{iqbal2019bibliometric}, the authors proposed keyword co-occurrence networks to detect closely-related research topics in computer networking. For their analysis, they considered papers published by top venues in this research area. Palchykov et al. \cite{palchykov2021network} proposed the network of scientific concepts and analyzed the structural features of this network, such as node density and degree distribution. They based their study on arXiv manuscripts and concepts elicited from the ScienceWISE.info platform. The authors of \cite{lafia2022subdivisions} constructed the network of research fields using co-cited datasets from different fields. They compiled a social science data archive for this study and demonstrated the inclusion of non-social science fields, such as information systems, artificial intelligence, and image processing, within this network. Cunningham et al. \cite{cunningham2022author} analyzed the multidisciplinarity of fields of study among papers related to network science and the COVID-19 pandemic. They constructed their field of study networks based on a heterogeneous network of papers, authors, and fields.

In 2012, ACM proposed a new version of the classification system for computing \cite{rous2012major}. Cassel and Buzydlowski \cite{cassel2020exploring} utilized this classification to investigate research patterns in computing from 1951 to 2017. They conducted time series analysis to assess how these patterns have evolved over time. In \cite{ullah2022analyzing}, the authors explored the co-authorship network of the Journal of Universal Computer Science to quantify interdisciplinary collaboration between ACM categories. They constructed a directed graph of computer science fields based on the aforementioned network.

Building upon prior research, we narrow our focus to the fields and subfields of computer science, aiming to illuminate their interactions and evaluate their interdisciplinary and multidisciplinary characteristics within the research landscape. Our study utilizes the ACM computing classification system as a standardized framework to investigate the well-defined fields and subfields of computer science. To the best of our knowledge, this is the first paper to analyze the relationships between research fields in computer science using a comprehensive dataset of different CS conferences, providing novel insights into their interconnections. Moreover, we emphasize that our study incorporates more recent data compared to previous works, allowing us to delve into the latest trends and advancements in computer science research.

Our contributions can be summarized as follows:

\begin{enumerate}
\item We investigate the distribution of computer science fields and subfields within the papers published by ACM.
\item We build the CS fields network and calculate centrality measures for its nodes, including betweenness centrality, to identify the most interdisciplinary and multidisciplinary research fields.
\item We examine the relationships between computer science subfields by analyzing the structural characteristics of the CS subfields network.
\item We discuss various features within the identified communities of the CS subfields network, including community size, prominent central nodes, and the most frequently occurring ACM fields.
\end{enumerate}

The remainder of this paper is organized as follows. In Section \ref{sec2} section, we explain the data collection process, as well as the methods and measures utilized in this work. In Section \ref{sec3}, we illustrate our discoveries and provide an analysis and interpretation of the obtained results. Finally, in Section \ref{sec4}, we provide a summary of the paper and explore potential avenues for future research.

\section{Methodology}\label{sec2}
In this section, we provide an overview of the data collected for this study. We describe the process of gathering the data and proceed to define the network constructed based on the extracted information. Subsequently, we outline the methods employed and the metrics calculated in our analysis.

\subsection{Dataset}
To collect the necessary data, we initially identified the top ACM conferences in various research areas, utilizing the ranking provided by Research.com. Through this process, we identified a total of nine conferences, which are listed in Table 1.

\begin{table}[h]
\caption{The best ACM conferences at different CS fields sorted by ranking.}\label{tab1}%
\begin{tabular}{c}
\toprule
Conference Name\\
\midrule
Human Factors in Computing Systems\\
Computer and Communications Security\\
Knowledge Discovery and Data Mining\\
International Conference on Management of Data\\
International Symposium on Computer Architecture\\
Foundations of Software Engineering\\
Symposium on the Theory of Computing\\
International Conference on Multimedia Retrieval\\
Symposium on Applied Computing\\
\botrule
\end{tabular}
\end{table}

From the proceedings of each conference in 2022, we gathered information including the titles and computer science fields and subfields of the research articles. The ACM digital library automatically classified the fields and subfields of 98.95\% (2083 out of 2095) of these papers. For the remaining 12 papers, we relied on the classification data provided by the authors in their manuscripts, while the remaining papers either lacked classification information or were not available online. An example of the conference data is displayed in Table \ref{tab2}.

\begin{table}[h]
\caption{An example of the conference data.}\label{tab2}%
\begin{tabular}{ccc}
\toprule
Conference & CS Fields & Paper\\
\midrule
CHI22 & \begin{tabular}{@{}c@{}}Security and privacy/ \\ Human-centered
computing\end{tabular} & 
\begin{tabular}{@{}c@{}}Understanding Privacy Switching Behaviour on \\ 
Twitter\end{tabular}\\
KDD22 & 
\begin{tabular}{@{}c@{}}Computing methodologies/ \\ Mathematics
of computing\end{tabular} & 
\begin{tabular}{@{}c@{}}TARNet: Task-Aware Reconstruction for \\ 
Time-Series Transformer\end{tabular}\\
\botrule
\end{tabular}
\end{table}

Furthermore, we generated individual datasets for each computer science field, consisting of the papers categorized within that particular field along with their respective subfields. It is important to mention that the classification of CS fields and subfields in our data aligns with the ACM computing classification system. An example of the field data is shown in Table \ref{tab3}.

\begin{table}[h]
\caption{An example of the field data (information systems).}\label{tab3}%
\begin{tabular}{ccc}
\toprule
Conference & CS Subfields & Paper\\
\midrule
CHI22 & \begin{tabular}{@{}c@{}}Information retrieval/ Information \\ systems applications\end{tabular} & 
What is Your Current Mindset?\\
KDD22 & 
\begin{tabular}{@{}c@{}}Information retrieval/World \\ Wide Web\end{tabular} & 
\begin{tabular}{@{}c@{}}Surrogate for Long-Term User Experience in \\ 
Recommender Systems\end{tabular}\\
\botrule
\end{tabular}
\end{table}

\subsection{CS fields network}
We constructed an undirected weighted network that visualizes the relationships between computer science fields based on papers published by ACM. In this network, each node represents a computer science field, and an edge with weight $W$ connecting fields $u$ and $v$ indicates that there are $W$ papers that belong to both fields $u$ and $v$.

\subsection{CS subfields network}
We constructed an undirected weighted network that represents the relationships between computer science subfields based on papers published by ACM. Each node in this network represents a computer science subfield, and an edge with weight $W$ connecting subfields $u$ and $v$ indicates that there are $W$ papers that belong to both subfields $u$ and $v$.

\subsection{Disparity filter}
We utilize the Disparity filter algorithm to decrease the network density by eliminating insignificant ties, ensuring the preservation of its multi-scale structure \cite{serrano2009extracting}. We utilize this technique to eliminate inconsequential edges within the CS fields network.

\subsection{Louvain community detection}
The Louvain algorithm for community detection optimizes the modularity of a network to identify communities by maximizing the difference between expected and observed connections \cite{blondel2008fast}. In our study, we employ this greedy approach to uncover communities within the CS fields network.

\subsection{Leiden community detection}
The Leiden community detection algorithm is an enhanced version of the Louvain algorithm that employs a fast local move approach. This algorithm refines the partitions iteratively to guarantee the connectedness of the identified communities. It exhibits improved computational efficiency and generates more precise partitions in comparison with the Louvain algorithm \cite{traag2019louvain}. We use this algorithm to specify communities within the CS fields network.

\subsection{Betweenness centrality}
The betweenness centrality of each node in a network measures the frequency with which a node lies on short paths between other pairs of nodes. The formula for calculating betweenness centrality is as follows:

\begin{eqnarray}
    B_u = \sum_{ij}{\frac{\sigma(i, u, j)}{\sigma(i, j)}},
\end{eqnarray}

Where $\sigma(i, u, j)$ represents the number of shortest paths between nodes $i$ and $j$ that pass through node or edge $u$, and $\sigma(i, j)$ represents the total number of shortest paths between $i$ and $j$. The summation is carried out over all pairs $i$ and $j$ of distinct nodes \cite{freeman2002centrality}. We employ this metric to determine the level of interdisciplinarity among different fields and subfields of computer science.

\subsection{Weighted degree centrality}
The weighted degree centrality of each node in a network is determined by the sum of the weights of the edges connected to that node. The formula for calculating weighted degree centrality is as follows:

\begin{eqnarray}
    WD(u) = \sum_{v}{w(v, u)},
\end{eqnarray}

where $v$ is a neighbor of $u$, and $w(v, u)$ represents the weight of the edge connecting $v$ and $u$ \cite{wei2012degree}. We utilize this measure to assess the degree of multidisciplinarity among various subfields of computer science.

\section{Results and discussion}\label{sec3}
In this section, we present a detailed analysis of our findings and offer insights into their interpretations.

\subsection{Exploring CS subfields}
In this part, we analyze the distribution of CS subfields to identify the most popular research areas within each field of computer science. For this purpose, we only consider the papers for which the subfields are specified.

\begin{figure}[ht]
\centering
  \includegraphics[width=11cm,
  keepaspectratio]{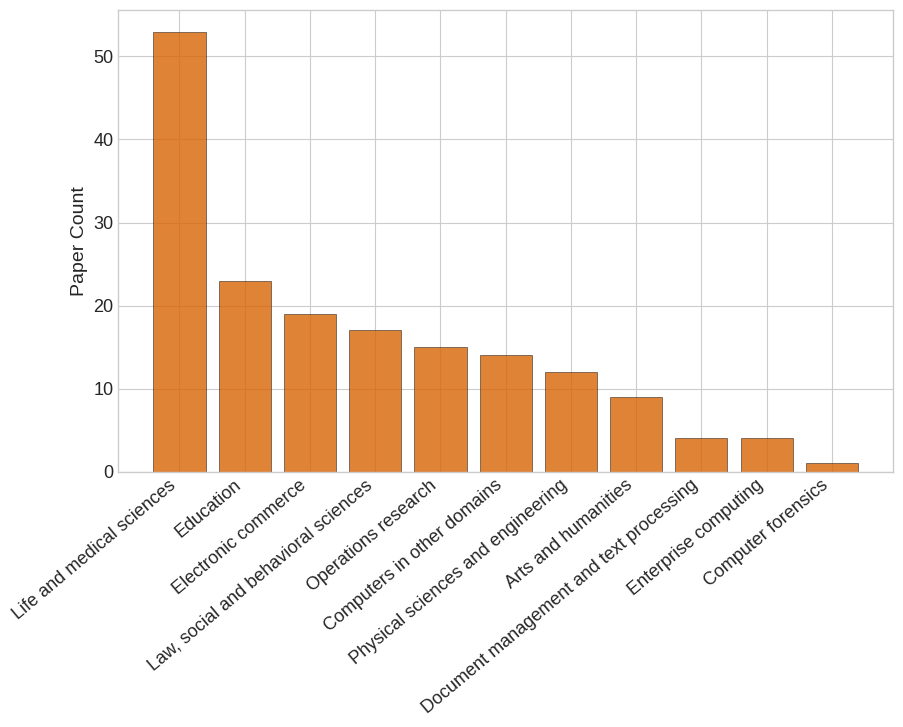}
\caption{Bar plot of applied computing subfields.}
\label{fig1}
\end{figure}

Figure \ref{fig1} illustrates the frequency of appearance of subfields within applied computing among ACM papers. It shows that life and medical sciences is the most common subfield in this area, followed by education and electronic commerce. These findings support previous studies highlighting the significance and prevalence of bioinformatics, education, and e-commerce as interdisciplinary research topics in CS \cite{luscombe2001bioinformatics, pears2002describing, ngai2002literature}.

\FloatBarrier

\begin{figure}[ht]
\centering
  \includegraphics[width=12cm,
  keepaspectratio]{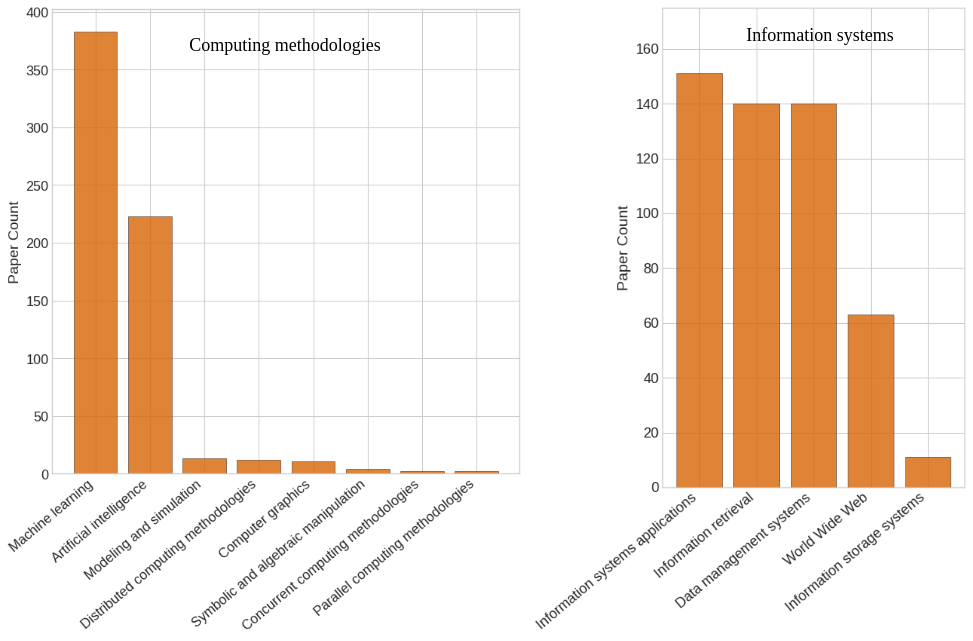}
\caption{Bar plots of computing methodologies and information systems subfields.}
\label{fig2}
\end{figure}

Figure \ref{fig2} presents the distribution of papers across different subareas of computing methodologies and information systems. It reveals that machine learning (ML) and artificial intelligence (AI) have the highest numbers of papers. This observation aligns with the findings of Xu et al. \cite{xu2021artificial}, who suggest that many computer scientists utilize AI and ML techniques in their scholarly work. Additionally, Figure \ref{fig2} shows that the majority of research articles in the field of information systems focus on the applications of these systems, confirming the findings of \cite{abu2009utilizing} that information systems are applicable to various domains.

\FloatBarrier

\begin{figure}[ht]
\centering
  \includegraphics[width=12cm,
  keepaspectratio]{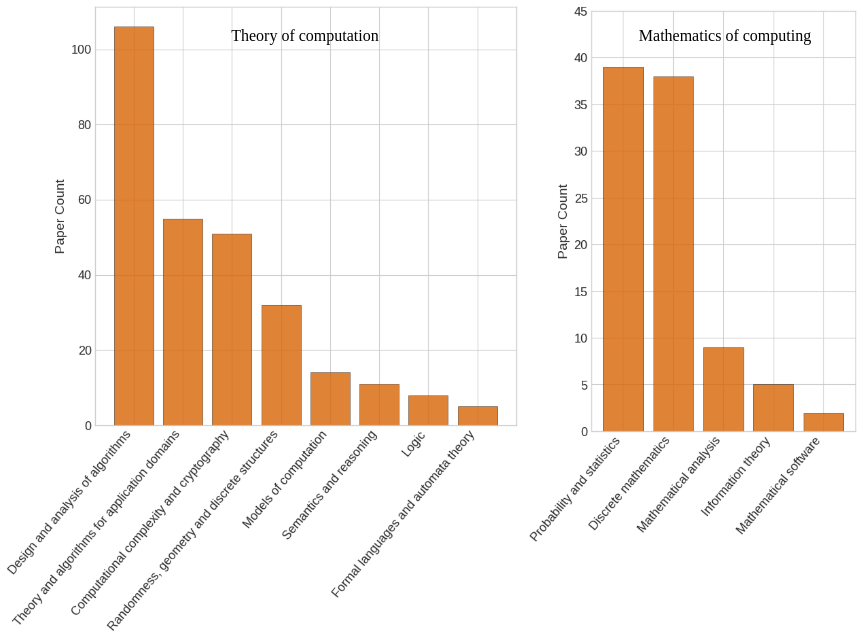}
\caption{Bar plots of theory of computation and mathematics of computing subfields.}
\label{fig3}
\end{figure}

Figure \ref{fig3} displays the distribution of papers among subfields of theory of computation and mathematics of computing. The design and analysis of algorithms has the highest frequency of research articles in the field of theory of computation, consistent with previous studies \cite{Shaltiel2004} highlighting it as the most common research theme among theoretical computer scientists. Moreover, probability and statistics, along with discrete mathematics, are prominent subfields. This result is expected since probability and statistics serve as critical foundations for machine learning \cite{el2015machine}, and algorithms are built upon concepts from discrete mathematics \cite{goodrich2001algorithm}.

\FloatBarrier

\begin{figure}[ht]
\centering
  \includegraphics[width=10cm,
  keepaspectratio]{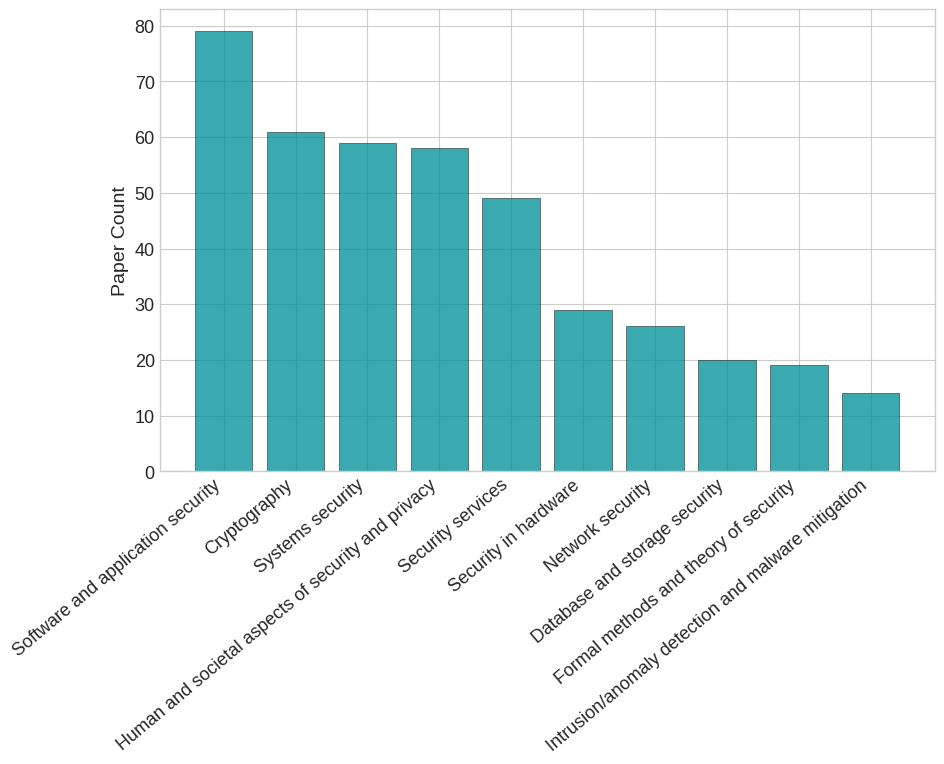}
\caption{Bar plot of security and privacy subfields.}
\label{fig4}
\end{figure}

Figure \ref{fig4} showcases the distribution of papers among subfields of security and privacy. Software and application security emerges as the dominant subarea, which is understandable given that security and privacy play integral roles in software systems \cite{ganji2015conflicts}.

\FloatBarrier

\begin{figure}[ht]
\centering
  \includegraphics[width=10cm,
  keepaspectratio]{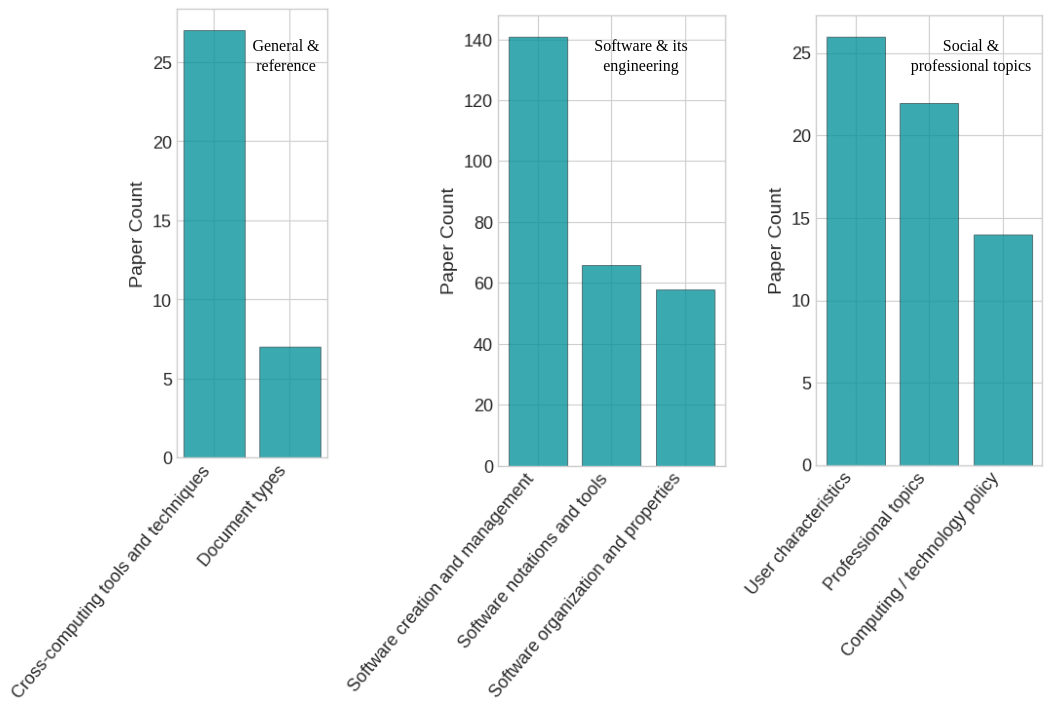}
\caption{Bar plots of general and reference, software and its engineering, and social and professional topics subfields.}
\label{fig5}
\end{figure}

Figure \ref{fig5} presents the distribution of subfields within general and reference, software and its engineering, and social and professional topics. Among general and reference subfields, cross-computing tools and techniques has the highest frequency. In software and its engineering, software creation and management stands out as the prominent research area, while user characteristics dominates the subfields of social and professional topics.

\FloatBarrier

\begin{figure}[ht]
\centering
  \includegraphics[width=6cm,
  keepaspectratio]{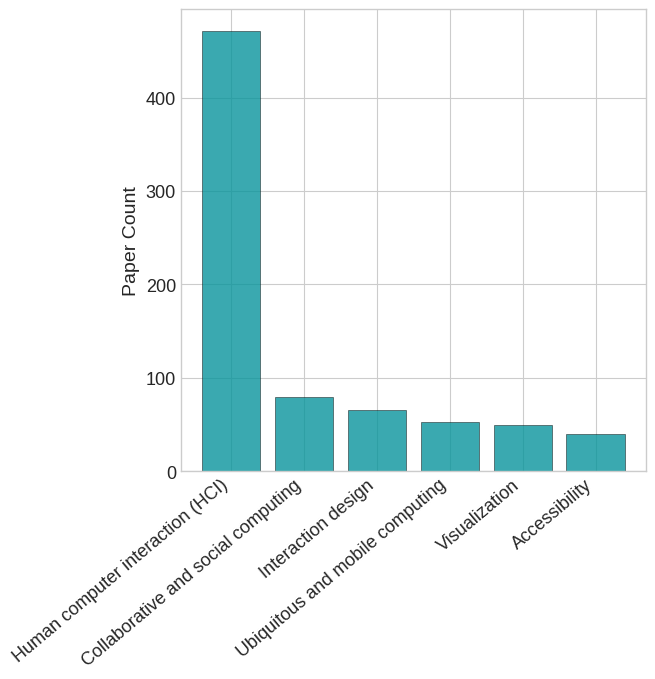}
\caption{Bar plot of human-centered computing subfields.}
\label{fig6}
\end{figure}

Figure \ref{fig6} depicts the bar plot of subfields within human-centered computing. It shows that human-computer interaction (HCI) has the highest frequency, aligning with previous studies \cite{kamppuri2006expanding} highlighting HCI as the dominant research area among all subfields of computer science.

\FloatBarrier

\begin{figure}[ht]
\centering
  \includegraphics[width=9cm,
  keepaspectratio]{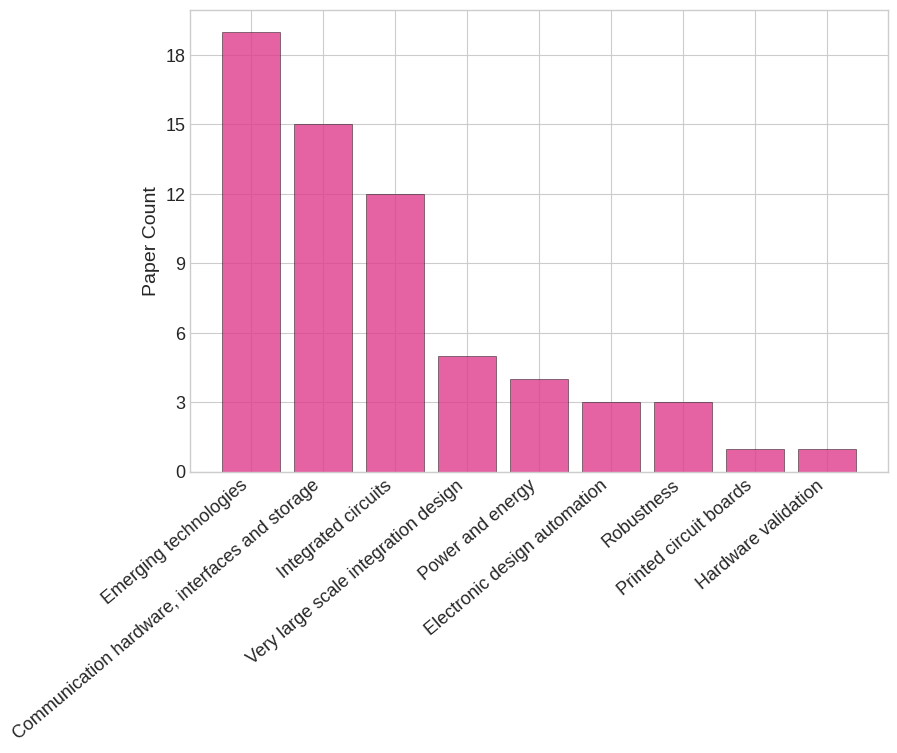}
\caption{Bar plot of hardware subfields.}
\label{fig7}
\end{figure}

Based on the distribution presented in Figure \ref{fig7}, emerging technologies is identified as the most prevalent subfield within hardware. These hardware technologies find wide applications in other subfields of CS, such as machine learning \cite{schneiderhardware, sze2017hardware}.

\FloatBarrier

\begin{figure}[ht]
\centering
  \includegraphics[width=8cm,
  keepaspectratio]{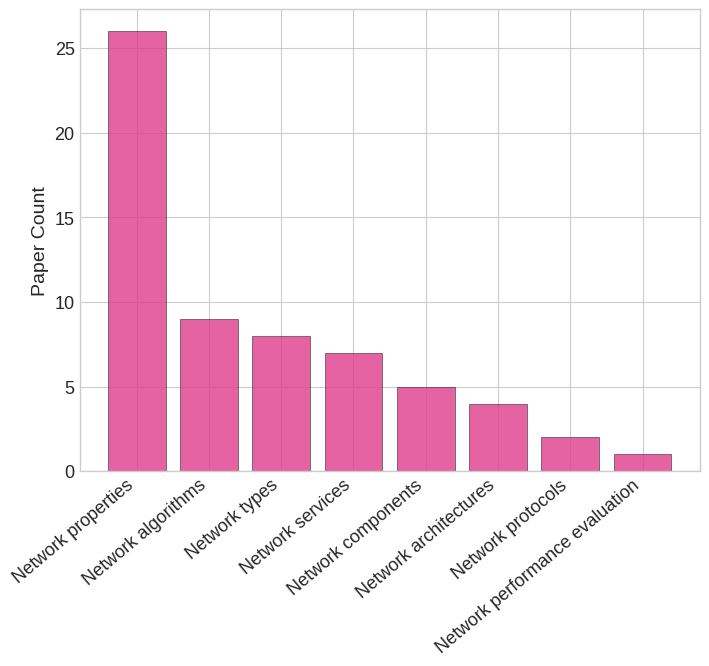}
\caption{Bar plot of networks subfields.}
\label{fig8}
\end{figure}

Figure \ref{fig8} presents the distribution of subfields within networks subfields. Network properties has the highest frequency, while other subfields within networks subfields have fewer than 10 papers.

\FloatBarrier

\begin{figure}[ht]
\centering
  \includegraphics[width=5cm,
  keepaspectratio]{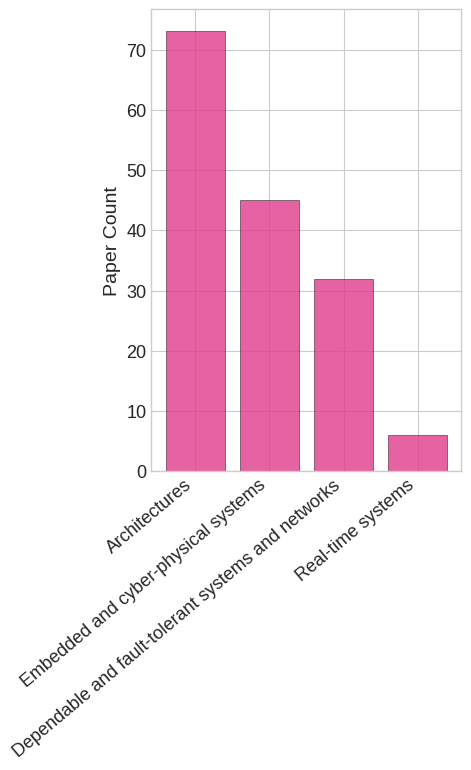}
\caption{Bar plot of computer systems organization subfields.}
\label{fig9}
\end{figure}

Figure \ref{fig9} showcases the distribution of subfields within computer systems organization. Architectures emerges as the most common research theme in this field of computer science, encompassing various types of computer architectures, including serial, parallel, and distributed architectures.

\FloatBarrier

\subsection{Analyzing CS fields network}
In this part, we delve into the CS fields network to gain deeper insights into the interconnections among various research fields in computer science. To filter out insignificant edges, we employ the disparity filter algorithm \cite{serrano2009extracting}, resulting in a 24.62\% reduction in network density (from 0.987 to 0.744). Next, we calculate the betweenness centrality of each research area to identify the most interdisciplinary field in computer science. Figure 10 demonstrates that the highest value of this metric is observed in the computing methodologies and security and privacy subfields. This finding is consistent with Aggarwal's research \cite{article}, which emphasizes the broad applicability of artificial intelligence, a prominent subfield within computing methodologies, across various research domains. Additionally, our previous study \cite{kalhor2023diversity} revealed that computing methodologies is the prevailing field of interest among computer science faculty members at top North American universities. Moreover, Landau \cite{landau2008privacy} highlights the extensive scope of security and privacy, encompassing various areas within science, technology, engineering, and mathematics (STEM). Thus, we can infer that research in the area of security and privacy has increased among computer science scholars in the last few years compared to previous decades \cite{cassel2020exploring}.

In addition to betweenness centrality, we analyze the frequency of each field among the research articles. Notably, human-centered computing emerges as the field with the highest number of papers. This finding aligns with the study conducted by Wania and colleagues \cite{wania2006mapping}, which highlights that human-computer interaction (HCI) is a multidisciplinary area that bridges various fields of computer science. Computing methodologies, on the other hand, secures the second rank and is recognized as the most productive research field \cite{fiala2017computer}.

\begin{figure}[ht]
\centering
  \includegraphics[width=12cm,
  keepaspectratio]{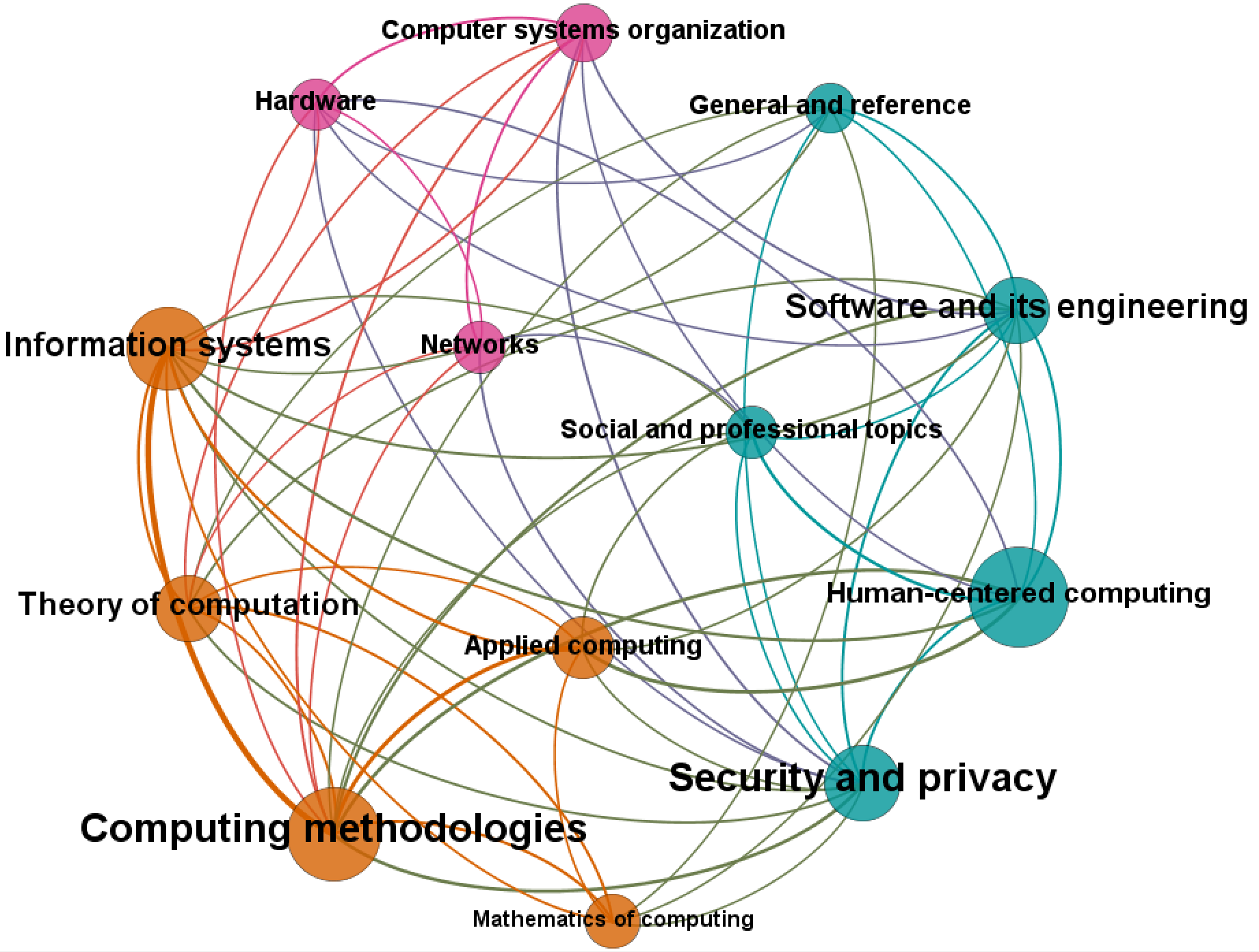}
\caption{The CS fields network. Each node's color indicates its community membership. The size of a node corresponds to the frequency of its associated field among papers, while the size of its label reflects its betweenness centrality. The thickness of an edge indicates its weight.}
\label{fig10}
\end{figure}

We further utilize the Louvain community detection algorithm \cite{blondel2008fast} to identify the communities within the CS fields network. As depicted in Figure \ref{fig10}, we observe three distinct communities, characterized by their respective colors (orange, blue, and pink), which can be interpreted as representing theory-based, software-based, and hardware-based fields, respectively.

\FloatBarrier

\subsection{Assessing CS subfields network}
\begin{figure}[ht]
\centering
  \includegraphics[width=14cm,
  keepaspectratio]{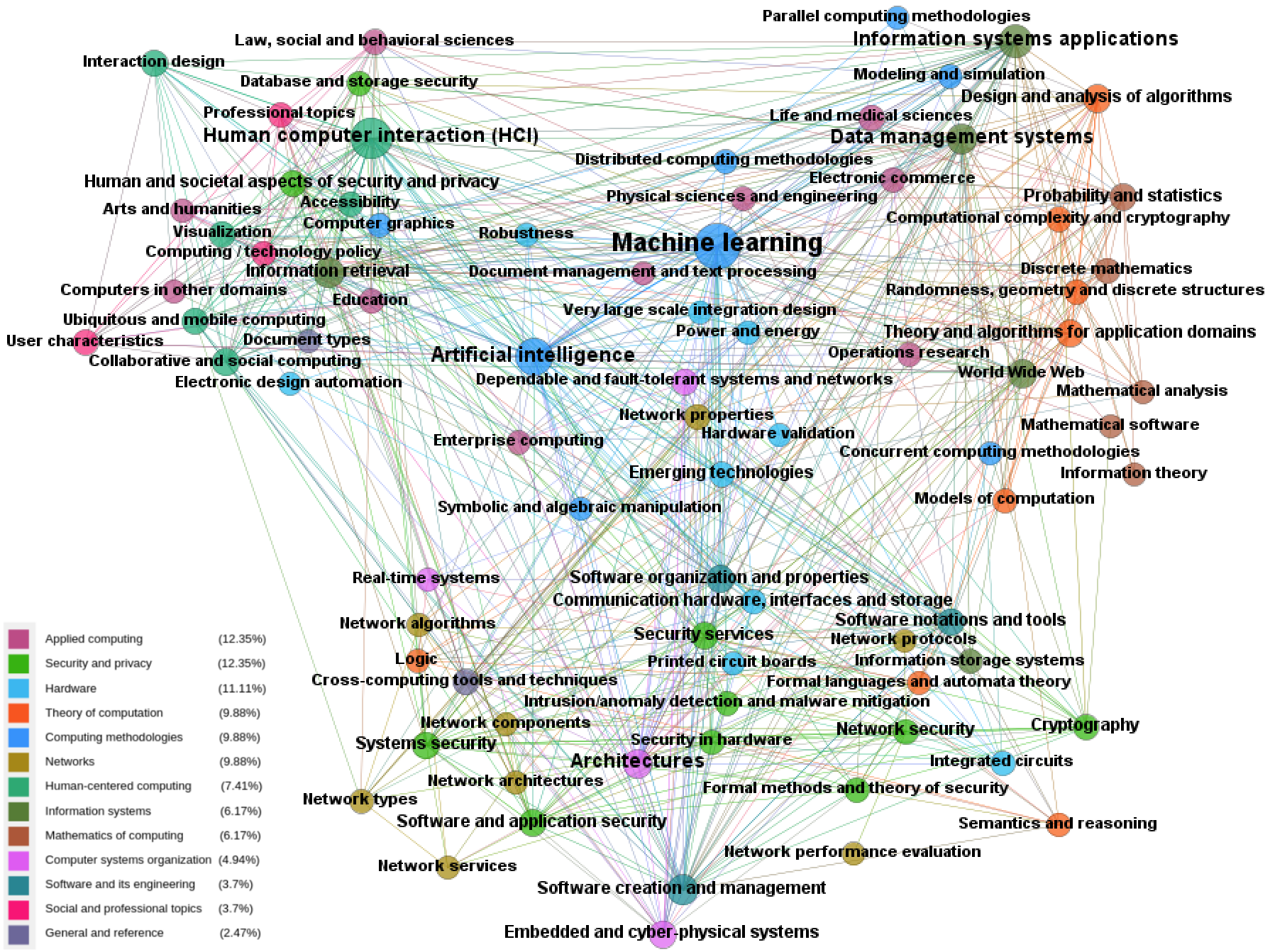}
\caption{The CS subfields network. Each node is color-coded according to its parent field. The size of a node reflects its weighted degree centrality, while the size of its label represents its betweenness centrality. The thickness of an edge signifies its weight.}
\label{fig11}
\end{figure}

In this section, we analyze the CS subfields network to examine the relationships between computer science subfields in the scholarly domain. The network comprises 81 subfields connected by 640 weighted edges. It is a connected network with a density of 0.1975. We calculate centrality metrics, including weighted degree and betweenness centralities, to assess the multidisciplinarity and interdisciplinarity levels of the nodes, respectively. Figure \ref{fig11} provides an overview of the CS subfields network. The subfields with the highest values of weighted degree centrality are ML, HCI, and AI, respectively. These findings align with previous studies \cite{suto2016comparison, wania2006mapping, arencibia2022evolutionary}, suggesting that these subfields are highly multidisciplinary within computer science. Furthermore, based on the betweenness centrality values, the most interdisciplinary subfields are ML, AI, and architectures, in that order. This finding is consistent with prior research \cite{athmaja2017survey, whittaker2018ai}.

\FloatBarrier

\begin{figure}[ht]
\centering
  \includegraphics[width=14cm,
  keepaspectratio]{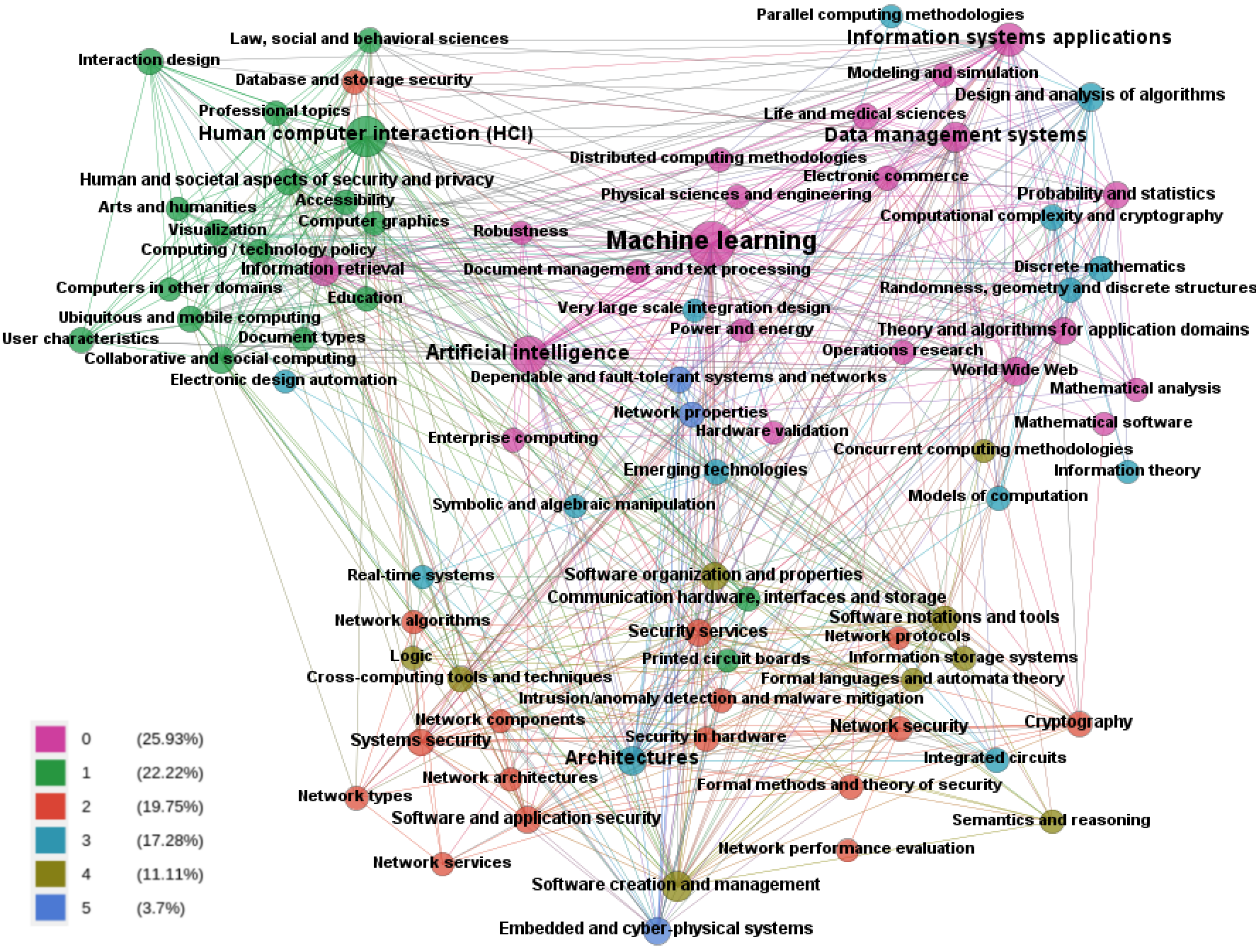}
\caption{The CS subfields network, with communities detected via the Louvain algorithm.}
\label{fig12}
\end{figure}

Figure \ref{fig12} presents the CS subfields network with the same configuration, but with nodes colored based on the communities identified by the Louvain community detection algorithm. The visualization shows six distinct communities that have been detected. The statistical information of these communities is presented in Table \ref{tab4}, which includes details such as community size, density, the most central fields based on betweenness centrality, and the most frequent computer science fields within each community.

Analyzing Table \ref{tab4}, we observe that theory of computation emerges as the most common CS field in two different communities. This finding supports the notion that theory of computation has broad applications in both hardware and software engineering disciplines \cite{bqa2019pedagogical, hannay2007systematic}.

Furthermore, according to the table, community 0 emerges as the largest community identified by the Louvain algorithm, predominantly comprising applied computing subfields. Within this community, machine learning takes the spotlight as the most central node, which aligns with the extensive utilization of machine learning techniques across diverse research domains that extend beyond computer science \cite{angra2017machine}. Additionally, community 5 exhibits the highest density and encompasses three fields primarily focused on computer systems organization, with embedded and cyber-physical systems occupying a central position. Moving on to community 1, it stands as the second largest community, where HCI takes center stage as the most central node. Furthermore, the parent field of HCI, human-centered computing, emerges as the most common field within this community.

\FloatBarrier

\begin{table}[h]
\caption{Statistics of the Louvain communities within the CS subfields network.}\label{tab4}%
\begin{tabular}{ccccc}
\toprule
\multicolumn{1}{c}{} & Size & Density & Most Central Subfield & ACM Fields\\
\midrule
0 & 21 & 0.4429 & Machine learning & Applied computing\\
1 & 18 & 0.4575 & 
\begin{tabular}{@{}c@{}}Human computer \\ interaction (HCI)\end{tabular} &
Human-centered computing\\
2 &
16 &
0.4083 &
Network security &
Security and privacy\\
3 &
14 &
0.2967 &
Architectures &
\begin{tabular}{@{}c@{}}Hardware, Theory of \\ computation\end{tabular} \\
4 &
9 &
0.5833 &
\begin{tabular}{@{}c@{}}Software creation and \\ management\end{tabular} &
\begin{tabular}{@{}c@{}}Software and its engineering, \\ Theory of computation\end{tabular} \\
5 &
3 &
1.0 &
\begin{tabular}{@{}c@{}}Embedded and \\ cyber-physical systems\end{tabular} &
\begin{tabular}{@{}c@{}}Computer systems \\ organization\end{tabular} \\
\botrule
\end{tabular}
\end{table}
\FloatBarrier

\begin{figure}[ht]
\centering
  \includegraphics[width=14cm,
  keepaspectratio]{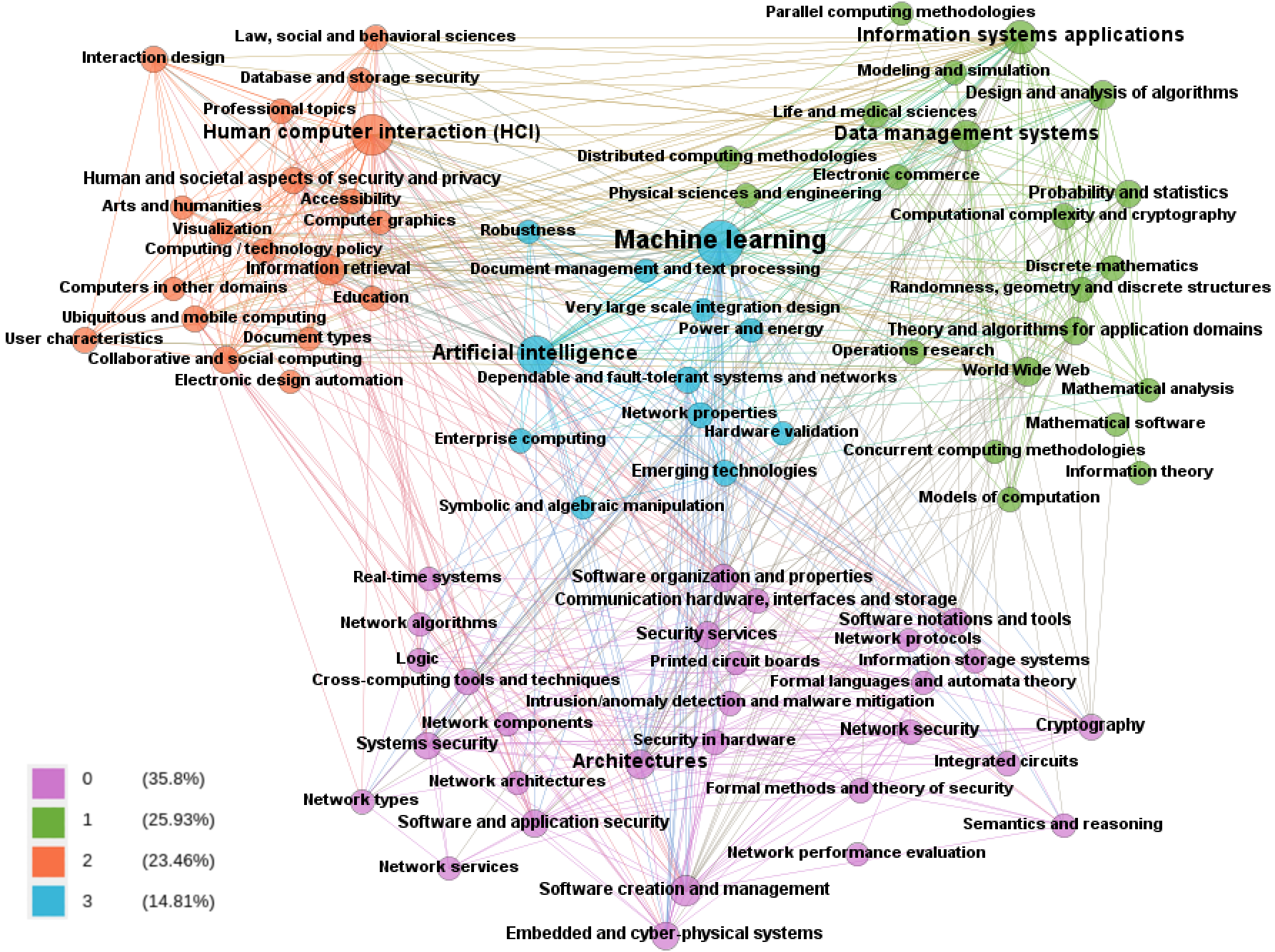}
\caption{The CS subfields network, with communities detected via the Leiden algorithm.}
\label{fig13}
\end{figure}

Figure \ref{fig13} depicts the CS subfields network with nodes colored according to the communities identified by the Leiden community detection algorithm. Four Leiden communities are detected within this network. An overview of statistical features of these communities are provided in Table \ref{tab5}. The largest community consists mostly of security and privacy subfields, and its most central node is architectures. The highest density is for community 3 with hardware as the prevailing field and machine learning as the most central node. The second largest community (community 1) has data management systems as the most central subfield and mathematics of computing and theory of computation as most frequent fields.

\FloatBarrier

\begin{table}[h]
\caption{Statistics of the Leiden communities within the CS subfields network}\label{tab5}%
\begin{tabular}{ccccc}
\toprule
\multicolumn{1}{c}{} & Size & Density & Most Central Subfield & ACM Fields\\
\midrule
0 & 29 & 0.3399 & Architectures & Security and privacy\\
1 & 21 & 0.3667 & Data management systems &
\begin{tabular}{@{}c@{}}Mathematics of computing, \\ Theory of computation\end{tabular} \\
2 &
19 &
0.4737 &
\begin{tabular}{@{}c@{}}Human computer interaction \\ (HCI)\end{tabular} &
Human-centered computing\\
3 &
12 &
0.5606 &
Machine learning &
Hardware \\
\botrule
\end{tabular}
\end{table}
\FloatBarrier

\section{Conclusion}\label{sec4}
In this study, we conducted a comprehensive analysis of computer science fields and subfields using the latest ACM conference papers in this research domain. Our investigation encompassed an assessment of the distribution of subfields within each field, revealing HCI, ML, and AI as the dominant subfields among the research articles. Furthermore, we examined the CS fields network by calculating betweenness centrality values and analyzing the frequency of its nodes. Our analysis highlighted computing methodologies and privacy and security as the most interdisciplinary fields, as indicated by their highest betweenness centrality scores. It is noteworthy that human-centered computing exhibited the highest frequency among the papers.

Additionally, we employed the Louvain community detection algorithm to classify nodes into three groups: theory-based, software-based, and hardware-based fields. Furthermore, we constructed the CS subfields network to explore the interconnectedness among various computer science subfields. Based on weighted degree and betweenness centrality values, we identified machine learning as the most multidisciplinary and interdisciplinary subfield within computer science. Moreover, we applied both the Louvain and Leiden community detection algorithms to uncover communities within the network and determined the most central subfields and the most frequent fields within these communities. Interestingly, the theory of computation emerged as the most frequent field in two distinct Louvain communities, while architectures occupied a central position in the largest Leiden community.

The findings of this study have practical implications for computer science departments in universities, as they can leverage these insights to create interdisciplinary research opportunities for students in fields like ML and HCI. By facilitating collaboration with researchers from other disciplines, universities can enhance the quality of research conducted by their students substantially.

\backmatter

%\bmhead{Supplementary information}

\section*{Declarations}
\subsection*{Funding}
No funding was received for conducting this study.

\subsection*{Competing interests}
The authors declare that they have no competing interests.

\subsection*{Ethics approval} 
Not applicable.

\subsection*{Consent to participate}
Not applicable.

\subsection*{Consent for publication}
Not applicable.

\subsection*{Availability of data and materials}
The dataset generated and analyzed during the current study is available in the \textit{CS Fields \& Subfields Data} repository, \href{https://github.com/kalhorghazal/CS-Fields-Subfields-Data}{https://github.com/kalhorghazal/CS-Fields-Subfields-Data}.

\subsection*{Authors' contributions}
All authors read and approved the final manuscript.

%\begin{itemize}
%\item Funding
%\item Conflict of interest/Competing interests (check journal-specific guidelines for which heading to use)
%\item Ethics approval 
%\item Consent to participate
%\item Consent for publication
%\item Availability of data and materials
%\item Code availability 
%\item Authors' contributions
%\end{itemize}

%\noindent
%If any of the sections are not relevant to your manuscript, please include the heading and write `Not applicable' for that section. 

%%===================================================%%
%% For presentation purpose, we have included        %%
%% \bigskip command. please ignore this.             %%
%%===================================================%%
\bibliography{sn-bibliography}% common bib file
%% if required, the content of .bbl file can be included here once bbl is generated
%%\input sn-article.bbl

\end{document}